\documentclass[10pt,twocolumn]{article}

\usepackage{graphicx}
\usepackage{dblfloatfix}   

\usepackage[font=small,labelfont=bf,labelsep=period]{caption}

\usepackage{titlesec}
\titleformat{\section}
  {\normalfont\Large\bfseries\raggedright\hyphenpenalty=10000\exhyphenpenalty=10000}
  {\thesection}{0.75em}{}

\titleformat{\subsection}
  {\normalfont\normalsize\bfseries\raggedright\hyphenpenalty=10000\exhyphenpenalty=10000}
  {\thesubsection}{0.75em}{}

\titlespacing*{\section}{0pt}{2.0ex plus .2ex minus .2ex}{2ex}
\titlespacing*{\subsection}{0pt}{1.0ex plus .2ex minus .2ex}{2.0ex}

\usepackage[a4paper,margin=1in]{geometry}
\usepackage{multirow}
\usepackage{amsmath,amssymb,amsfonts,amsthm,mathrsfs,bm}
\usepackage[title]{appendix}
\usepackage{xcolor}
\usepackage{textcomp}
\usepackage{manyfoot}
\usepackage{booktabs}
\usepackage{algorithm}
\usepackage{algorithmicx}
\usepackage{algpseudocode}
\usepackage{listings}
\usepackage[numbers,sort&compress]{natbib}
\usepackage[colorlinks=true,allcolors=blue]{hyperref}

\theoremstyle{plain}

\theoremstyle{definition}

\theoremstyle{remark}

\begin{document}

\twocolumn[
\begin{center}
{\LARGE\bfseries
A plug-and-play superconducting quantum controller at millikelvin temperatures enables exceeding 99.9\% average gate fidelity
\par}

\vspace{1em}

{\normalsize
Kuang Liu$^{1}$, Zhiyuan Wang$^{1,2}$, Xiaoliang He$^{1}$, Siqi Li$^{1,2}$, Hao Wu$^{1,2}$, Xiangyu Ren$^{1,2}$,\\
Zhengqi Niu$^{1}$, Wanpeng Gao$^{1}$, Chenluo Zhang$^{1,2}$, Pei Huang$^{1,2}$, Yu Wu$^{1,2}$,\\
Liliang Ying$^{1,2}$, Wei Peng$^{1,2}$, Jaw-Shen Tsai$^{3,4}$, Zhi-Rong Lin$^{1,2}$*
\par}

\vspace{0.5em}

{\small
$^{1}$State Key Laboratory of Materials for Integrated Circuits, Shanghai Institute of Microsystem and Information Technology, Chinese Academy of Sciences, Shanghai 200050, China\par
$^{2}$University of Chinese Academy of Sciences, Beijing 100049, China\par
$^{3}$Graduate School of Science, Tokyo University of Science, Shinjuku, Tokyo 162-8601, Japan\par
$^{4}$RIKEN Center for Quantum Computing (RQC), Wako-shi, Saitama 351-0198, Japan\par
\vspace{0.3em}
*Corresponding author. E-mail:  \href{mailto:zrlin@mail.sim.ac.cn}{zrlin@mail.sim.ac.cn}
}
\end{center}

\vspace{0.8em}

\noindent\textbf{Abstract}\par
\noindent
The development of large-scale superconducting quantum computing requires efficient in-situ control methods that allow high-fidelity operations at millikelvin temperatures. Superconducting circuits based on Josephson junctions offer a promising solution due to their high speed, low power dissipation, and cryogenic nature. Here, we report a superconducting quantum controller that enables direct chip-to-chip interconnection with qubits at 10 mK and high-fidelity, all-digital manipulation. Randomized benchmarking reveals a uniformly high average Clifford fidelity of 99.9$\%$ with leakage to high energy levels on the order of 10\textsuperscript{-4}, and an estimated average gate operation energy of 0.121 fJ, demonstrating the potential to resolve the control bottleneck in superconducting quantum computing. 

\vspace{1.5em}
]

\section{Introduction}\label{sec1}
Superconducting qubits are a promising platform for building a universal quantum computer \cite{kjaergaard2020superconducting}, and recent demonstrations have achieved hundred-qubit processors with gate fidelities above the surface code threshold for quantum error correction (QEC) \cite{arute2019quantum,gong2021quantum,ball2021first}. Nevertheless, scaling superconducting qubits to the millions for practical fault-tolerant quantum computing demands further advances in efficient control schemes \cite{gambetta2017building,gidney2021factor}. The cryogenic nature of superconducting qubits requires high-fidelity operations at millikelvin (mK) temperatures, and therefore, demands cross-temperature signal interconnections that introduce wiring complexity, control overhead, and heat-load constraints, limiting the maximum scale to thousands in existing cryostats \cite{mohseni2024build,franke2019rent}.

Various cryogenic integrated control approaches have been explored to resolve the scalability challenge, including complementary metal-oxide-semiconductor (CMOS) electronics \cite{bardin201929,pauka2021cryogenic}, photonic link \cite{lecocq2021control}, and superconducting circuits \cite{mcdermott2018quantum,li2019hardware,jokar2022digiq,PRXQuantum.3.010350,doi:10.1063/5.0083972,9586326}. Among these, superconducting circuits are promising candidates for simultaneously realizing high-fidelity manipulation at mK temperatures, owing to their low power dissipation, high operational speed, and inherent compatibility with superconducting qubits. However, the superconducting circuits control scheme cannot yet match the performance of state-of-the-art room-temperature electronics, as phonon-mediated \cite{PhysRevApplied.11.014009, PhysRevB.108.064512} and photon-assisted \cite{PRXQuantum.4.030310,PhysRevLett.132.017001} quasiparticle poisoning, along with high-energy-level leakage, degrade gate fidelity, undermining the algorithm complexity achievable in fault-tolerant quantum computing \cite{fowler2012surface,gidney2021factor,google2025quantum}.

\begin{figure*}[th]
\centering
\includegraphics[width=1.0\textwidth]{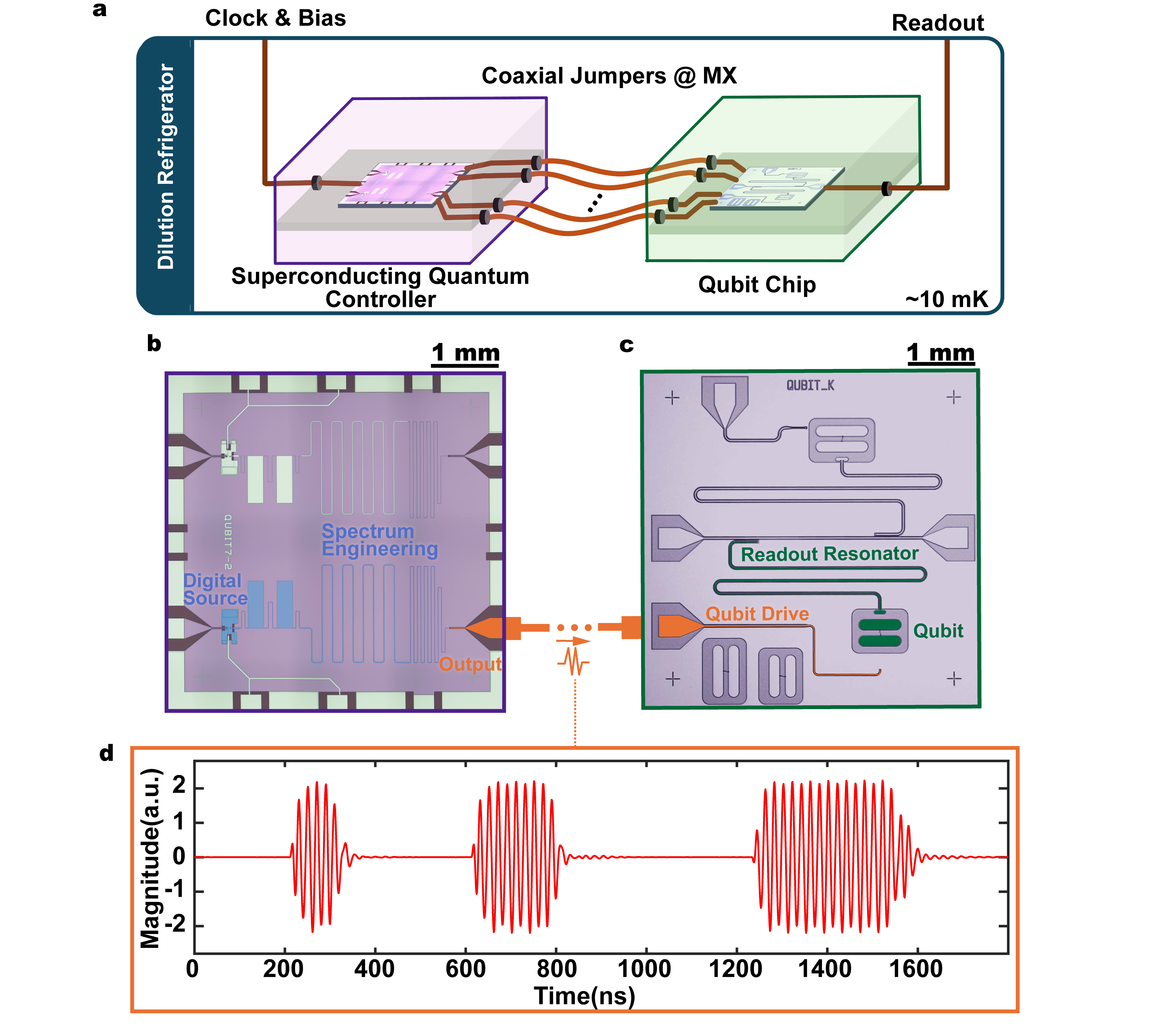}%
\caption{\label{fig1} \textbf{Superconducting quantum controller for high-fidelity qubit control at 10 mK.} 
\textbf{a,} Scheme of the proposed superconducting quantum controller for qubit manipulation at cryogenic temperatures. The controller is chip-to-chip integrated with qubits at the 10 mK mixing chamber stage, and the generated digital control signals are transported via coaxial cables. 
\textbf{b, c,} Micrograph of superconducting quantum controller chip (b) and superconducting qubit chip (c). 
\textbf{d,} Representative output of the superconducting quantum controller obtained by heterodyne demodulation referenced 50~MHz below 4.8 Ghz (2$\omega_\mathrm{SQC}$).}
\end{figure*}

In this article, we report a plug-and-play superconducting quantum controller for high-fidelity all-digital qubit manipulation at millikelvin temperatures. It suppresses quasiparticle poisoning and leakage to high energy levels, generating clean, power-efficient digital control signals, and supports direct chip-to-chip interconnection with qubits at millikelvin temperatures, minimizing wiring complexity and losses while enhancing system flexibility and scalability (Fig. \ref{fig1}a). The chip is fabricated by depositing niobium (Nb) layers on high-resistivity silicon via the established CMOS nanofabrication process, enabling rapid iteration and production. Randomized benchmarking shows that the superconducting quantum controller, integrated with a qubit chip at 10 mK, achieves a uniformly high average Clifford fidelity of 99.9$\%$ with leakage to higher-energy levels on the order of $10^{-4}$, approaching the decoherence limit and surpassing the quantum error correction threshold. The estimated average gate operation energy is 0.121 fJ, validating the proposed approach as a high-fidelity, scalable solution to the control challenge in superconducting quantum systems.

\section{Device Architecture and Characteristic at Millikelvin Temperatures}\label{sec2}

\begin{figure*}[t]
\centering
\includegraphics[width=1.0\textwidth]{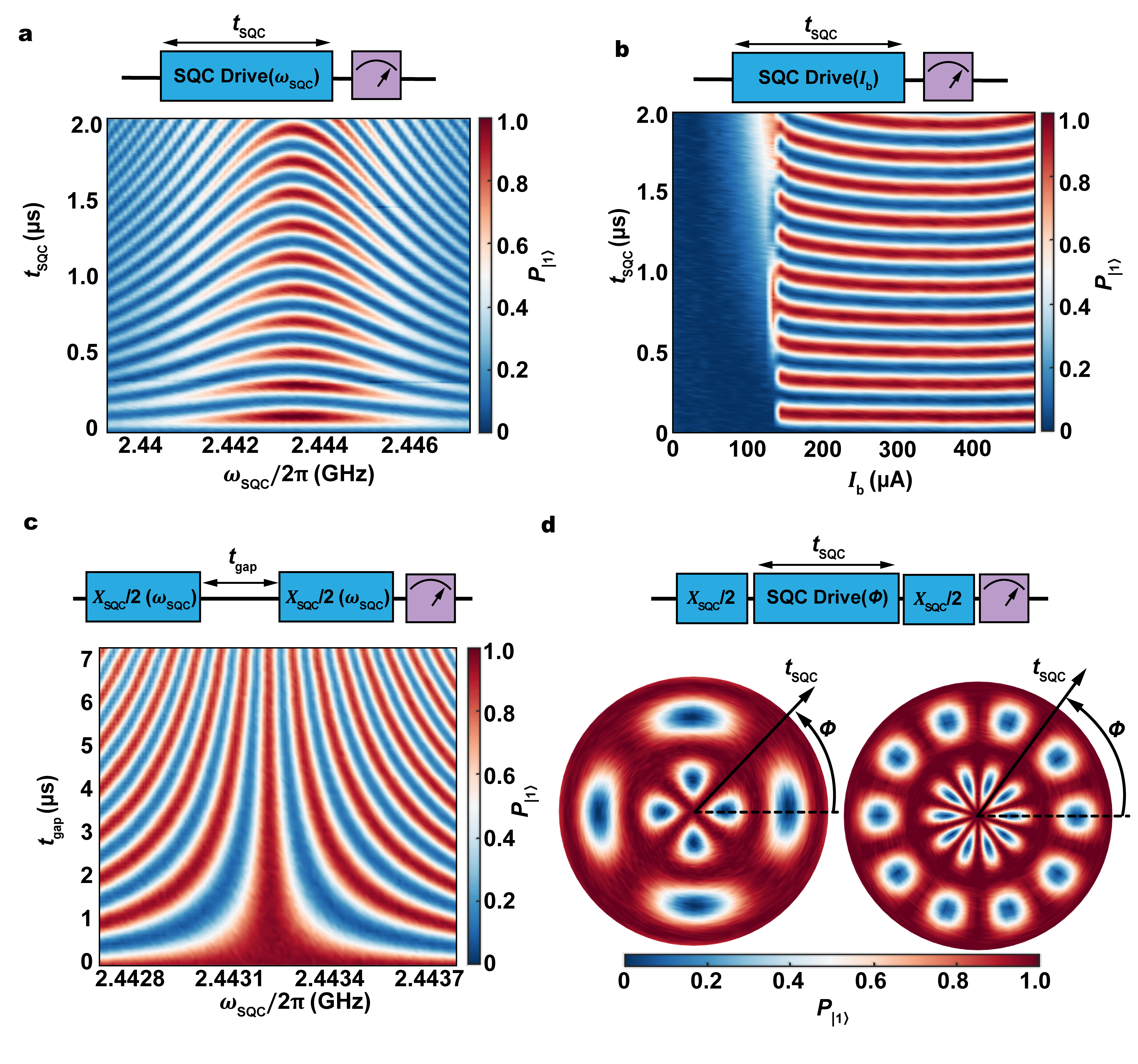}
\caption{\label{fig2}  \textbf{Single qubit control with a superconducting quantum controller.}
\textbf{a,} Rabi oscillation as a function of $\omega_{\mathrm{SQC}}$.
\textbf{b,} Rabi oscillation as a function of $I_\mathrm{b}$ with $\omega_{\mathrm{SQC}}$ fixed to ${2\pi}\cdot 2.4432\; \mathrm{GHz}$. The clear chevron pattern indicates the control signal is robust and has a sharp on/off transition.
\textbf{c,} Ramsey interference as a function of $\omega_{\mathrm{SQC}}$.
\textbf{d,} Orthogonal rotation by phase adjustment at subharmonic frequency ${\omega_{\mathrm{SQC}}}={\omega_{\mathrm{01}}}/2$ (left) and ${\omega_{\mathrm{SQC}}}={\omega_{\mathrm{01}}}/5$ (right).
}
\end{figure*}

The proposed superconducting quantum controller leverages a superconducting digital source to generate synchronized trains of single-flux-quantum (SFQ) voltage pulses for qubit drive. Each pulse equals a voltage-time integral of one flux quantum, a quantization protocol that immunizes the signal against noise. To alleviate quasiparticle poisoning, a fully passive superconducting bias network is employed to eliminate static power related quasiparticles, and a low-critical-current topology is adopted to minimize quasiparticles dynamically generated during active switching. As for high-energy-level leakage, an on-chip spectrum-engineering unit is integrated to minimize reflections, maximize power-transfer efficiency, and enhance fidelity by filtering out-of-band noise from SFQ pulses. The device was fabricated on a VHF-precleaned high-resistivity silicon substrate using a foundry-compatible multilayer Nb superconducting process. All features were defined by a wafer-scale, fully photolithographic patterning and etching flow identical to the established CMOS nanofabrication process, ensuring chip reproducibility. The superconducting quantum controller enables direct interconnection with qubits at mK temperatures via cryogenic RF connections, including wire bonding, coaxial jumpers, and interposers, in a plug-and-play manner that supports flexible chip deployment to optimize electromagnetic fields and suppress antenna parasitic coupling, thereby enhancing fidelity.

\section{Fidelity Benchmarking}\label{sec3}
\begin{figure*}[t]
\centering
\includegraphics[width=1.0\textwidth]{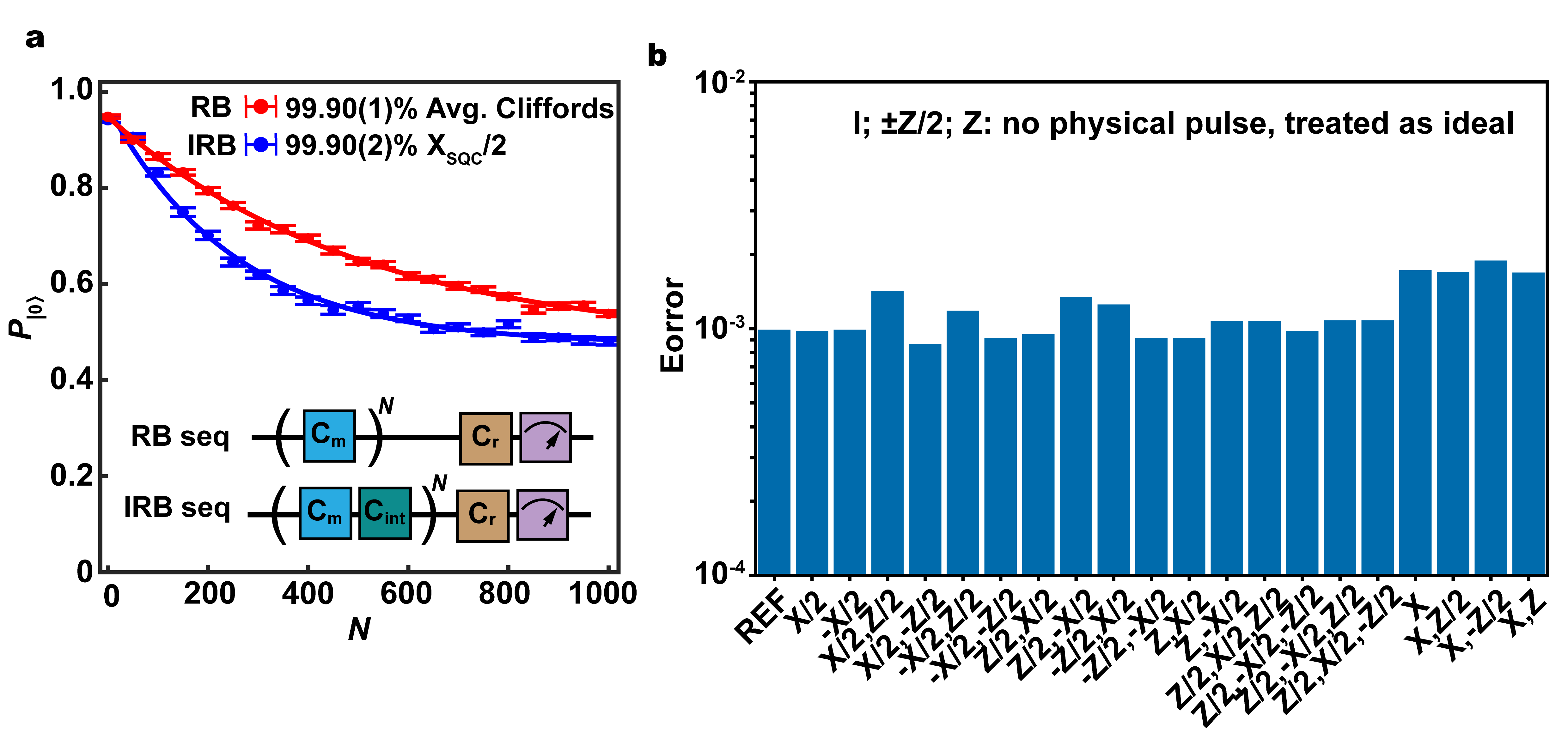}
\caption{\label{fig3}\textbf{Gate fidelity benchmarking.} 
\textbf{a,} Depolarizing curves for the reference RB sequence and the IRB sequence with $X_{\mathrm{SQC}}/2$ gates that demonstrate the relationship between ground state probability and the number of Clifford gates. Each data point is the average of 100 random sequences.
\textbf{b,} Average and individual gate fidelities excluding the identity, $\mathrm{\pm Z}$, $\mathrm{Z}$.
}
\end{figure*}

To characterize device performance at cryogenic temperatures, the superconducting quantum controller chip (Fig.~\ref{fig1}b) and transmon qubit chip (Fig.~\ref{fig1}c) connected via coaxial cables are placed in the mixing chamber stage of a dilution refrigerator with a 10 mK base temperature. The measured time response of enveloped SFQ pulse sequence generated by the controller with length $t_{\mathrm{SQC}}$ of 80, 160, and 300 ns, and a clock frequency $\omega_{\mathrm{SQC}}$  of $2\pi\cdot2.4$ GHz is shown in Figure~\ref{fig1} (d), demonstrating the ability to produce a flat-top enveloped signal with arbitrary duration for coherent qubit control at mK temperatures. Rabi experiments are performed to verify the drive coherence, and the qubit chevron pattern is obtained by sweeping $\omega_{\mathrm{SQC}}$ and $t_{\mathrm{SQC}}$ (Fig.~\ref{fig2}a). The $\omega_{\mathrm{SQC}}$ is set near half of the qubit’s $\lvert 0\rangle$ to $\lvert 1\rangle$  transition frequency $\omega_\mathrm{01}$ so that the SFQ pulses recur at a fixed phase of the qubit precession, ensuring each pulse produces a phase-coherent incremental rotation. The Rabi chevron pattern is extracted by sweeping the controller's bias current $I_\mathrm{b}$  while keeping $\omega_{\mathrm{SQC}}$ fixed (Fig.~\ref{fig2}b). The clear oscillation observed once  exceeds the device's operation threshold indicates that the control signal is robust and has a sharp on/off transition. The control parameters for $\pi$ and $\pi/2$ rotations, denoted $\mathrm{X} _\mathrm{SQC}$  and $\mathrm{X} _\mathrm{SQC}/2$, are also identified, with corresponding durations of 100 ns (244 SFQ pulses) and 50 ns (122 SFQ pulses), respectively. Ramsey experiments are then performed using a sequence of two $\mathrm{X} _\mathrm{SQC}/2$ gates separated by a variable free-evolution interval $t_\mathrm{gap}$, and interference fringes oscillating at a frequency of $\left|2(\omega_{\mathrm{SQC}}-\omega_\mathrm{01}/2) \right|$ are produced by sweeping $\omega_{\mathrm{SQC}}$ around $\omega_\mathrm{01}/2$ (Fig.~\ref{fig2}c). The orthogonal rotation is achieved by inserting an additional SFQ pulse train between two $\mathrm{X} _\mathrm{SQC}/2$ with a variable phase $\phi$ and duration $t_{\mathrm{SQC}}$, while keeping all other parameters unchanged (Fig.~\ref{fig2}d). When the control signal frequency $\omega_{\mathrm{SQC}}$ is near the subharmonic condition $\omega_\mathrm{01}/N$, the qubit accumulates a $2\pi N$ Z-phase within one clock cycle, resulting in the observed 2$N$-fold symmetry. The protocol provides a practical route to calibrate orthogonal single-qubit rotations and serves as the basis for a universal single-qubit gate set.

The single-qubit gates' fidelities are assessed using randomized benchmarking (RB) and interleaved randomized benchmarking (IRB). The complete set of 24 Clifford operations is implemented by combining calibrated $\pm \mathrm{X} _\mathrm{SQC}/2$ and $\mathrm{X} _\mathrm{SQC}$ with pulse-free virtual-Z ($\mathrm{\pm Z/2}$ and $\mathrm{Z}$) through phase-frame updates, minimizing the effective gate duration and improving fidelity. The principle of virtual-Z gates is that the qubit precesses by 4$\pi$ between pulses under subharmonic conditions $\omega_\mathrm{SQC}=\omega_\mathrm{01}/2$, so updating the clock phase performs a Z-frame update, where shifts of $\pm \pi/4$ and $\pi/2$ correspond to virtual $\mathrm{\pm Z/2}$ and $\mathrm{Z}$ gates, respectively. The Clifford set exhibits a uniform duration distribution, with most gates lasting $\pi/2$ while only four gates ($\mathrm{X}$, $\mathrm{X\cdot Z/2}$, $\mathrm{X\cdot -Z/2}$, and $\mathrm{X\cdot Z}$) require a $\pi$ length, resulting in uniformly high fidelities and low performance variation. Fig.~\ref{fig3}a shows depolarization curves from standard RB and IRB with an $\mathrm{X} _\mathrm{SQC}/2$  gate interleaved between random Clifford operations  with clock frequency ${\omega_{\mathrm{SQC}}}\approx 2\pi\cdot 2.4432~\mathrm{GHz}$. Fitting the reference RB data yields an average Clifford fidelity of 99.90(1)$\%$ with error rates several times lower than the publicly reported record \cite{bernhardt2025quantum}, while interleaving $\mathrm{X} _\mathrm{SQC}/2$  gates gives an interleaved-gate fidelity of 99.90(2)$\%$, demonstrating the primitive SFQ $\pi/2$ rotation is in the 99.9$\%$ fidelity regime (error~$10^{-3}$). The uniformity of the gate set is quantified by evaluating errors through performing analogous interleaved sequences for each Clifford operation, as illustrated in Fig.~\ref{fig3}b. The identity and pure-Z operations ($\mathrm{I}, \mathrm{\pm Z/2, Z}$) are regarded as ideal, for they are realized virtually. The fidelities of the pulsed Cliffords are tightly clustered around 99.9$\%$, demonstrating consistently high performance across the gate set and the potential to build a practical fault-tolerant quantum computer.

\begin{figure*}[t]
\centering
\includegraphics[width=1.0\textwidth]{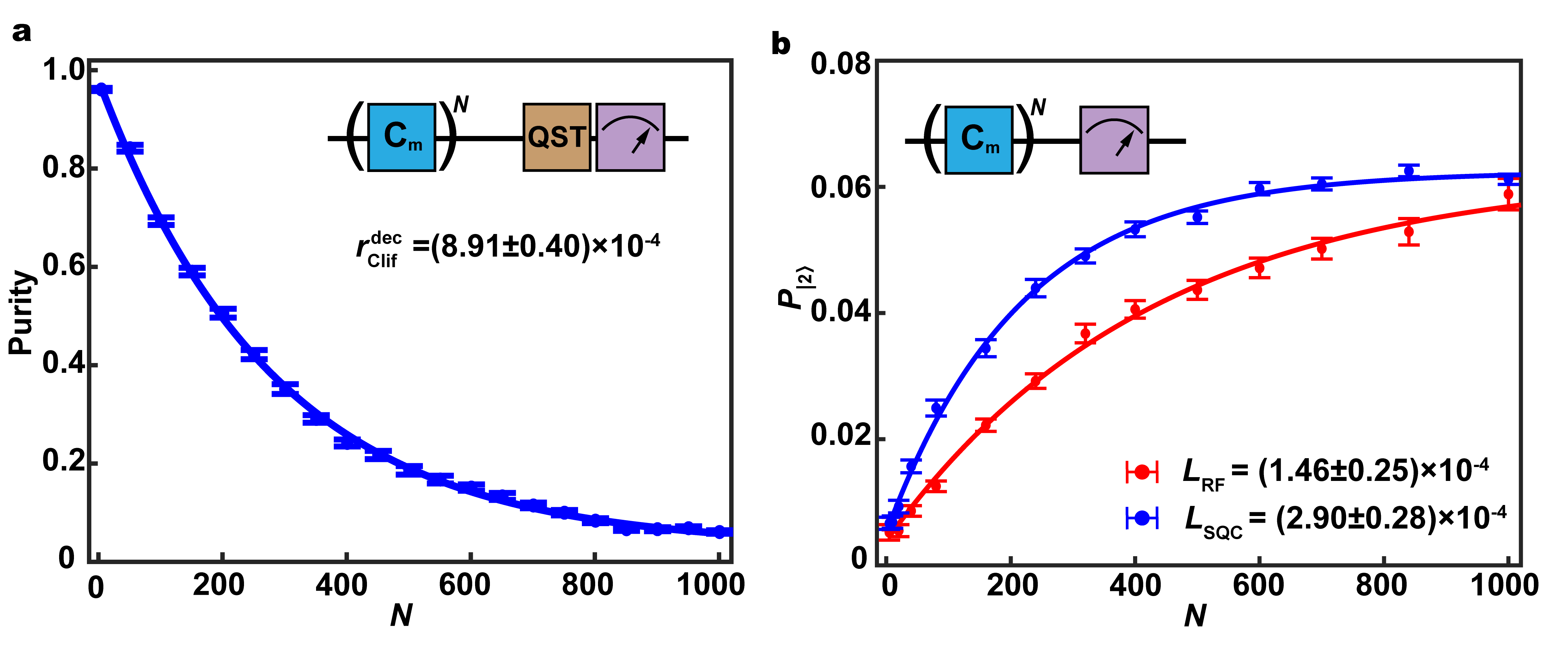}
\caption{\label{fig4} \textbf{Purity and leakage randomized benchmarking.} 
\textbf{a,} 
Purity RB by fitting the purity decay versus sequence length.
 \textbf{b,} Leakage RB by quantifying $\lvert 2\rangle$ state population after random Clifford sequences. Each data point is the average of 100 random sequences.}
\end{figure*}

The improvement can be attributed to the effective suppression of both quasiparticle poisoning and high-energy-level leakage. Nonequilibrium quasiparticles, which mainly propagate to superconducting quantum chips via substrate phonon transportation as well as high-frequency photon coupling and cause decoherence through Cooper-pair breaking, have been identified as one of the major factors degrading gate fidelity below its theoretical limit \cite{PhysRevApplied.11.014009, PRXQuantum.4.030310}. Although previous studies have explored flip-chip architectures to impede phonon-mediated propagation \cite{PRXQuantum.4.030310} and qubit-electrode geometry optimization to suppress photon-assisted coupling \cite{PhysRevLett.132.017001}, such strategies, while effective to some extent, fail to provide full-path blocking and undermine the controller's universality. The proposed superconducting quantum controller suppresses quasiparticle generation at the source through a fully passive superconducting bias network and a low-critical-current topology. Meanwhile, it employs a discrete architecture to impede substrate phonon propagation and electromagnetic-field shielding to block antenna-parasitic-coupling-induced photon transportation, forming a solid barrier against decoherence channels. High-energy-level leakage that induces undesired excitations is another major factor of gate error that is unresolved by the cryogenic control scheme \cite{10.1063/5.0304764}. The proposed control chip employs digital SFQ pulses as a noise-immune, robust qubit drive, and a spectrum-engineering unit to filter out high-frequency harmonics for waveform purification. All fidelity-enhancement methods of the superconducting quantum controller are self-contained, ensuring plug-and-play compatibility with various quantum chips while delivering superior overall performance.

\section{Purity and Leakage Analysis}\label{sec4}
To further quantify the error budget, purity randomized benchmarking (Purity RB) and leakage randomized benchmarking (Leakage RB) have been performed under the same Clifford gate set. Purity RB tracks the decay of state purity with sequence length and estimates the decoherence-limited contribution to gate error, complementing conventional RB that primarily reports an average depolarizing error \cite{PhysRevLett.117.260501,Wallman_2015}. Fig.~\ref{fig4}a demonstrates a decoherence-limited error per Clifford of $r_{Clif}^{dec}=8.91\times10^{-4}$, a value close to the average error extracted from standard RB in Fig.~\ref{fig3} (average Clifford fidelity~99.9$\%$), indicating that performance is approaching the coherence-imposed limit. Leakage RB has also been performed to quantify population transfer out of the computational subspace under long random sequences, as shown in Fig.~\ref{fig4}b \cite{PhysRevA.97.032306}. The assessments are conducted using both the pulses produced by a superconducting quantum controller and a conventional Gaussian enveloped microwave signal generated by room-temperature electronics for comparison, yielding leakage rates of $L_\mathrm{SQC}=2.90\times10^{-4}$ per Clifford and $L_\mathrm{RF}=1.46\times10^{-4}$ per Clifford, respectively. The modest difference between the two cases indicates that leakage from the controller remains strongly suppressed during randomized operation, reaching a low level comparable to those achieved with room-temperature electronics. The disparity may arise from off-resonant excitation caused by spectral residual in the SFQ drive sequence near the $\lvert 1\rangle$ to $\lvert 2\rangle$ transition frequency $\omega_{12}$ , which can therefore be addressed through spectrum engineering, such as pulse sequence optimization or quadrature compensation, approaches that have been validated as effective strategies to reduce leakage and attain high-fidelity gates within a short operation time \cite{liu2023single2,dtdk-kc2b,vozhakov2023speeding}.

\section{Energy Dissipation and Thermal Excitation Qualification}\label{sec5}

\begin{figure}[h]
\centering
\includegraphics[width=0.45\textwidth]{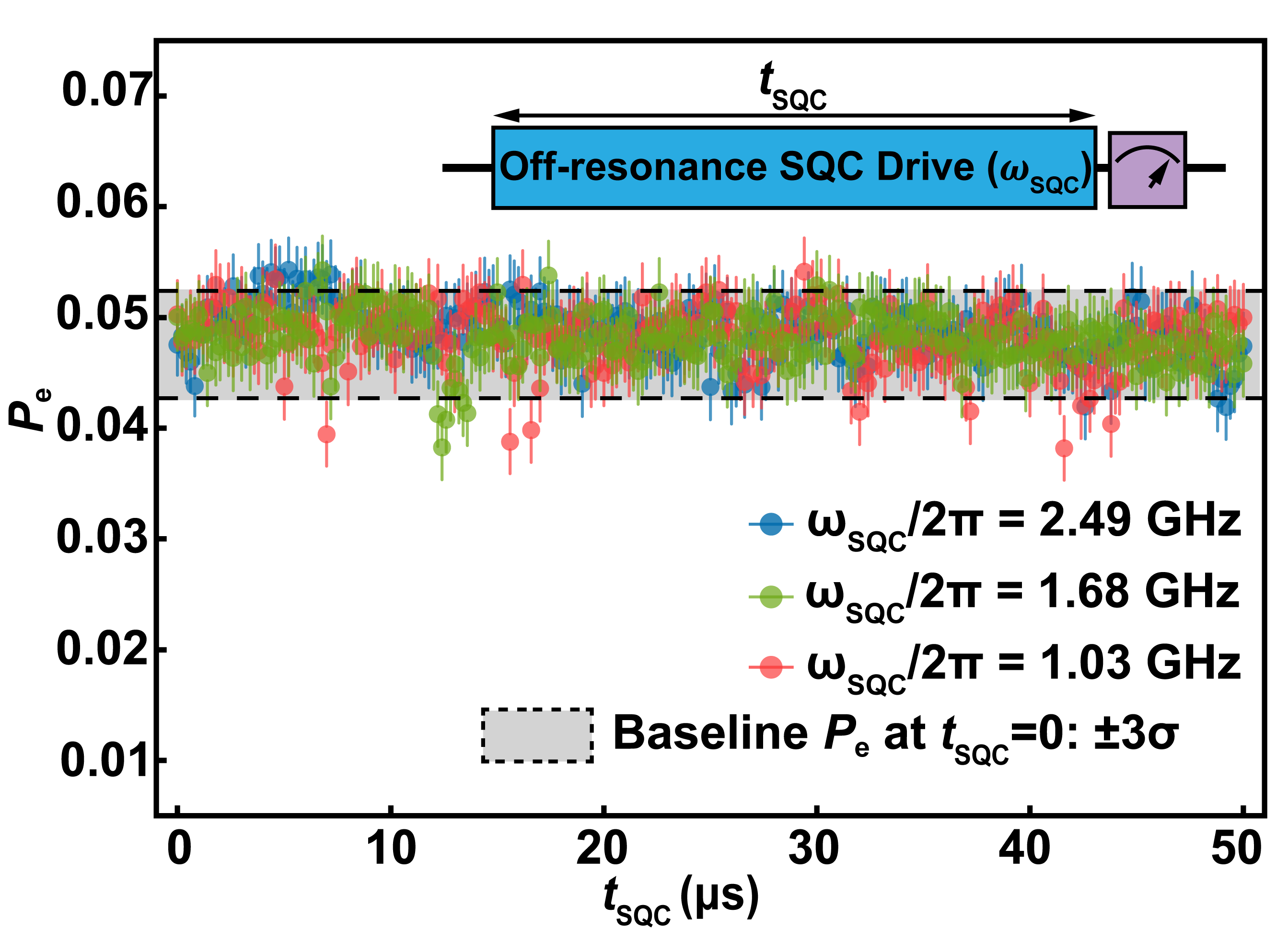}
\caption{\label{fig5} \textbf{Thermal excited-state population estimation.}
After the superconducting quantum controller operates for a duration $t_{\mathrm{SQC}}$  (0–50~$\mu$s) with clock frequencies detuned by 50 MHz from the subharmonic conditions: $\omega_{\mathrm{SQC}}\approx \omega_\mathrm{01}/2 + 2\pi\times 50~\mathrm{MHz}$ (blue), $\omega_\mathrm{01}/3 + 2\pi\times 50~\mathrm{MHz}$ (green), and $\omega_\mathrm{01}/5 + 2\pi\times 50~\mathrm{MHz}$ (red), the thermal excited-state population $P_e$ is extracted using JPA-assisted single-shot dispersive readout. The gray band denotes the $3\sigma$ interval obtained from 250 repeated baseline measurements when the controller is off. Obtained $P_e$ is independent of $t_{\mathrm{SQC}}$ and $\omega_{\mathrm{SQC}}$.}
\end{figure}

The total energy consumption is dominated by its dynamic contribution, given by $E=N\Phi_0 I_c$, as the fully passive superconducting bias network renders the static contribution negligible. Here, $N$ denotes the total number of SFQ pulses, with the 24 Clifford gates containing an average of 122 pulses each, yielding an estimated average gate operation energy consumption of 0.121 fJ. To probe controller-induced heating or nonequilibrium excitation, the qubit's thermal excited-state population $P_e$ was measured after activating the superconducting controller for variable durations. The controller operates for durations $t_\mathrm{SQC}=0\sim50\mu s$, with clock frequencies detuned by 50 MHz from the subharmonic conditions $\omega_\mathrm{01}/2$, $\omega_\mathrm{01}/3$ to avoid coherent driving. After each run, $P_e$ is extracted via single-shot dispersive readout aided by Josephson parametric amplifiers (JPA) and compared to a baseline distribution obtained from 250 repetitions when the controller is off (Fig.~\ref{fig5}). The baseline thermal population of the qubit chip is $P_e=(4.76\pm0.49)\%$ (mean ±3$\sigma$ over the 250 baseline repetitions), yielding an effective temperature $T_\mathrm{eff}=78.3\pm2.8~ \mathrm{mK}$ according to the thermal-equilibrium relation $P_e=\left [ 1+exp(\hbar\omega_\mathrm{01}/k_BT_\mathrm{eff} ) \right ]^{-1} $ with $\omega_\mathrm{01}/2\pi = 4.886 ~\mathrm{GHz}$. The negligible shift of $P_e$  beyond the 3$\sigma$ band indicates that any controller-induced thermal load, stray-photon excitation, or quasiparticle-assisted processes remain below the measurement sensitivity and do not raise the excited-state occupation, proving efficient operation at mK temperatures.

\section{Conclusion}\label{sec6}
We have reported a superconducting quantum controller supporting high-fidelity and all-digital qubit manipulation at millikelvin temperatures in a plug-and-play manner. The chip leverages the superconducting digital source along with the spectrum-engineering technique to efficiently generate a clean control signal, minimizing cross-temperature wiring complexity and losses while improving flexibility and scalability. Randomized benchmarking reveals a uniformly high average Clifford fidelity of 99.9$\%$, approaching the coherence-imposed limit, while the leakage to higher energy levels is on the order of $10^{-4}$, demonstrating suppression comparable to that of a conventional Gaussian enveloped microwave signal generated by room-temperature electronics. The estimated average gate operation energy is 0.121 fJ, and the distribution of qubit excited-state population validates efficient operation at mK temperatures.

Fidelity can be further improved by upgrading to a superconducting quantum controller with adjustable dual-pulse intervals \cite{liu2023single2}, which introduces an additional degree of freedom for envelope engineering and quadrature corrections to reduce leakage and unitary errors \cite{torosov2025optimization,dtdk-kc2b}. Moreover, black-box discrete search, especially genetic algorithms and learning-assisted deep exploration \cite{PhysRevApplied.6.024022,dalgaard2020global,bastrakova2023genetic,li2019hardware}, can be employed to optimize sequence duration for fast, high-fidelity operations. To complement the gate set, a two-qubit CZ gate will be realized by applying programmable driving signals to both the qubits and the coupler sequentially, yielding a low-crosstalk all-microwave implementation \cite{wang2023single,shirai2023all,gao2025all}. The scalability can be enhanced by developing superconducting tunable bandpass filters for microwave XY control and zero-static-power flux digital-to-analog converters for DC Z control, along with time- and frequency-division multiplexing architectures. These provide a viable route to address the control bottleneck and facilitate large-scale quantum computing.

\section*{Acknowledgements}
\addcontentsline{toc}{section}{Acknowledgements}
The authors acknowledge the use of the Superconducting Electronics Facility (SELF) at Shanghai Institute of Microsystem and Information Technology. This work is supported by the National Natural Science Foundation of China (Grants No. 92576207), the Strategic Priority Research Program of the Chinese Academy of Sciences (Grant No. XDB0670000), the Key-Area Research and Development Program of Guangdong Province, China (No. 2020B0303030002) and the Shanghai Science and Technology Program (No. 25LZ2600600).

\section*{Author Contributions}
\addcontentsline{toc}{section}{Author Contributions}
Z.L. conceived and supervised the project. Z.L. and K.L. planned the experiments. K.L. designed the superconducting quantum controller and the transmon qubit. K.L. fabricated the superconducting quantum controller sample, with contributions from W.P. and L.Y., who supported the fabrication process in the Superconducting Electronics Facility. X.H., H.W., X.R., Z.N., W.G., and Y.W. fabricated the transmon qubit sample. K.L., Z.W., and C.Z. performed the measurements. K.L., Z.W., and Z.L. analyzed the qubit-control performance and thermal-excitation data. P.H. fabricated the Josephson parametric amplifier used in the experiment. S.L., K.L., and Z.L. wrote the manuscript, with input from all authors.

\bigskip

\bibliographystyle{unsrtnat}
\bibliography{bibliography}

\end{document}